\documentclass[a4paper,11pt]{article}
\pdfoutput=1
cm \voffset = -2truecm

\usepackage{graphicx}
\usepackage{amsmath}
\usepackage{amssymb}
\usepackage{latexsym}
\usepackage[colorlinks]{hyperref}
\usepackage{color}
\usepackage{cite}
\usepackage{makeidx}
\usepackage{float}

\restylefloat{figure}
\newcommand{\bra}{\begin{array}}
\newcommand{\era}{\end{array}}
\newcommand{\beq}{\begin{equation}}
\newcommand{\eeq}{\end{equation}}
\newcommand{\bqr}{\begin{eqnarray}}
\newcommand{\eqr}{\end{eqnarray}}

\def\BC{\bb C}
\def\_\BC{\bbi C}

\def\no2 {{\textstyle{n\over 2}}}

\begin{document}
\begin{titlepage}
\setcounter{page}{1}
\renewcommand{\thefootnote}{\fnsymbol{footnote}}

\begin{flushright}
\end{flushright}

\begin{center}

{\Large \bf {Periodic Barrier Structure in AA-Stacked Bilayer
Graphene}}

\vspace{5mm}

{\bf Ilham Redouani}$^{a}$
and
{\bf Ahmed Jellal\footnote{\sf ajellal@ictp.it --
a.jellal@ucd.ac.ma}}$^{a,b}$

\vspace{5mm}

{$^{a}$\em Theoretical Physics Group,  
Faculty of Sciences, Choua\"ib Doukkali University},\\
{\em PO Box 20, 24000 El Jadida, Morocco}

{$^b$\em Saudi Center for Theoretical Physics, Dhahran, Saudi
Arabia}


\vspace{3cm}

\begin{abstract}
We study the charge carriers transport in an AA-stacked bilayer
graphene modulated by a lateral one-dimensional multibarrier
structure. We investigate the band structures of our system, that is made up of two shifted Dirac cones, for
finite and zero gap. We use the boundary conditions to explicitly
determine  the transmission probability of each individual cone
($\tau=\pm 1$) for single, double and finite periodic barrier
structure. We find that the Klein tunneling is only
possible when the band structure is gapless and can occur at
normal incidence as a result of the Dirac nature of the
quasiparticles. We observe that the band structure of the
{barriers} can have more than one Dirac points for
finite periodic barrier.
The
resonance peaks appear in the transmission probability, which
correspond to the positions of new cones index like
associated with $\tau=\pm 1$.
Two conductance
channels through different cones ($\tau=\pm 1$) are found
where 
the {total conductance}
has been studied and
compared to the cases of single layer and
AB-stacked bilayer graphene.

\end{abstract}
\end{center}

\vspace{3cm}


\end{titlepage}


\section{Introduction}

Graphene, single layer of pure carbon atoms in a honeycomb
lattice, has a gapless linear electronic spectrum near to the two
valleys of the first Brillouin zone
\cite{Novosolov,CastroNeto,Peres}. That yields  electrons in
graphene are described by a massless two dimensional relativistic
Dirac equation and the corresponding {electrons have
a chiral nature} \cite{Abergel,Ando}. This leads to many unusual
electronic properties and potential applications
\cite{CastroNeto,Peres}. Another interesting property is the Klein
tunneling \cite{Klein}, describing that the Dirac fermions can be
transmitted with probability one through a classically forbidden
region. Graphene can not only exist in the free state, but two or
more layers can stack above each other to form what is called
few-layer graphene, as example bilayer graphene is resulting
from a composition of two stacked sheets. 
Bilayer graphene has surged as another
{attractive} two-dimensional carbon material. In
addition, it is demonstrated that the bilayer graphene has a new
unusual physical properties and the spectrum is different from
that of single layer graphene. There are two dominant ways in
which the two layers can be stacked:  the AB-stacked
and AA-stacked bilayer graphene.

AB-stacked bilayer graphene or Bernal form is the basis of the
graphite from \cite{Bernal} and is usually derived. In which
only one atom in the lower layer lies directly below an other atom
in the upper layer and the other two atoms over the center of the
hexagon in the other layer. AB-stacked bilayer graphene has a gapless
quadratic dispersion relation, two conduction bands and two
valance bands, each pair is separated by an interlayer coupling
energy of order $\gamma_1=0.4\ eV$. Recent
{experiments} showed that the
AA-stacked bilayer graphene  could also exist and is a new stable stacking of
graphene \cite{Lee,deAndres,Liu,Ho,Borysiuk,Lobato}, where the A
sublattice  of the top layer is stacked directly above the same
sublattice of the bottom layer. In AA-stacked bilayer graphene, the energy
bands are just the double copies of single layer graphene bands
shifted up and down by the interlayer coupling $\gamma=0.2\ eV$. Additionally, it
has band structures that differ from those corresponding to single layer and
AB-stacked bilayer graphene.

Recently, many works have been reported in
order to investigate
the multibarrier structures in single layer graphene
\cite{Azarova,Park,Ghosh,Bliokh,Snyman,Nasir,Barbier,Bai}.
However, {much} less experimental and theoretical
works have been done on AB-stacked bilayer graphene
\cite{Bai,Barbier09}.
Motivated by different developments on the subject and in
particular \cite{Tabert,Sanderson,Azarova}, we investigate the
energy bands for finite and zero gaps in AA-stacked bilayer
graphene. 
Our theoretical model is based on the well established
tight binding Hamiltonian \cite{Tabert,Sanderson,Bena,Rakhmanov}
and we adopt the parametrization of the relevant intralayer and
interlayer couplings. We note that the electronic band structure
can be modified by the application of a periodic potential and the
gaps. We explicitly calculate the transmission probability of each individual cone
($\tau=\pm 1$) for single, double and finite periodic barrier
structure. We show that
the Klein tunneling is only
possible when the band structure is gapless and can occur at
normal incidence. This makes difference with respect to the case of
AB-stacked
bilayer graphene where
no
Klein tunneling ($T=1$) is expected 
\cite{Bai,Duppen}.
We show that
the existence of the intracone transition allows only two transmission channels in AA-stacked bilayer
graphene resulting in a total conductance, while the transmission in the intercone is zero. In
contrast, for AB-stacked bilayer graphene \cite{Duppen} we have intracone and intercone transitions, which are
possible between all four bands. In addition, the total conductance of AA-stacked bilayer graphene is
different with that of the single layer graphene \cite{Azarova}.

The rest of the paper is organized as follows. In section $2$, we
consider the multibarrier structure for the AA-stacked bilayer
graphene. We formulate our model by setting the Hamiltonian system
and calculating the associated energy bands in each potential
region. We obtain the spinor solution corresponding to each
regions. Using the transfer matrix at boundaries together with the
incident, transmitted and reflected currents we end up with
{two} transmission {probabilities}.
In section 3, we numerically present our results
of the transmission probability of each individual cone for
single, double and finite periodic barrier structure. In section
4, based on the obtained results 
we study two conductance channels through different cones and
as well as the corresponding total conductance. 
We conclude our work in the final section.

\section{Energy bands}

\begin{figure}[H]
 \centering
 \includegraphics[width=7.5cm, height=5.5cm]{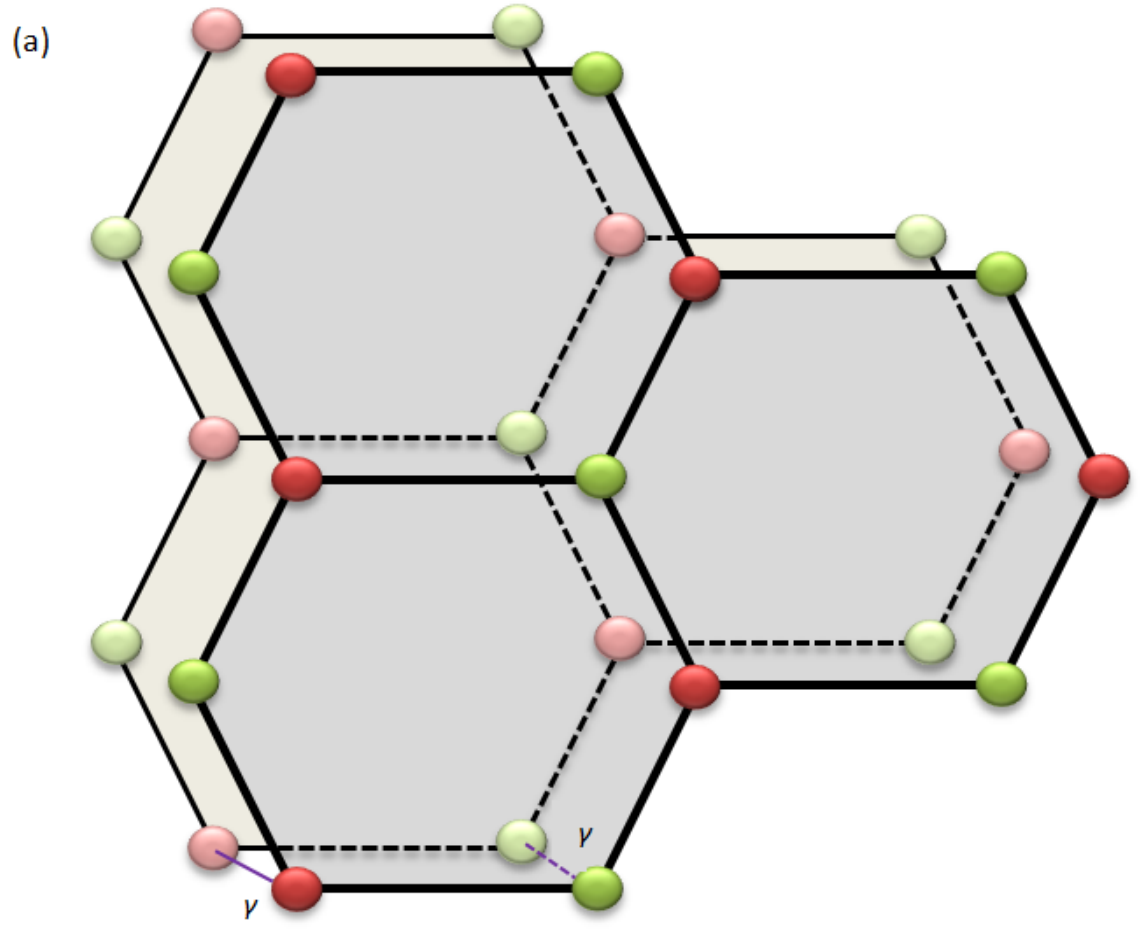}
 \ \ \ \includegraphics[width=3cm, height=2cm]{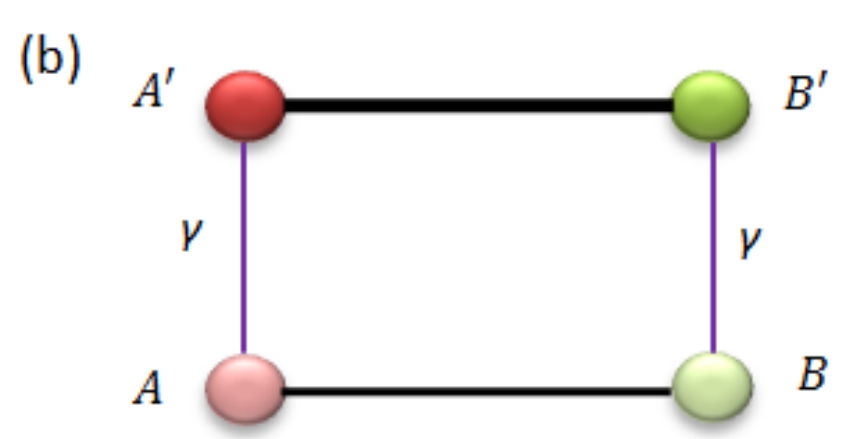}
 \\ \includegraphics[width=16.5cm, height=2.5cm]{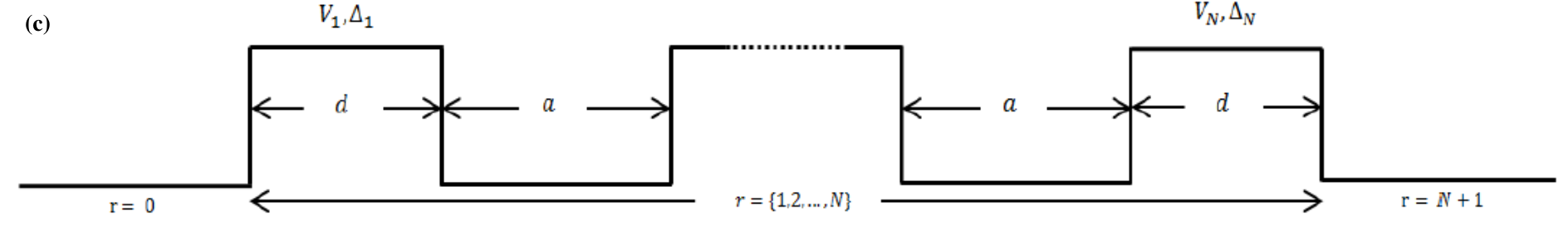}
 \caption{\sf(a): Illustration of lattice structure of
 AA-stacked bilayer graphene, which is consisting of two coupled single layer
 graphene. (b): unit cell of AA-stacked consists of four
 atoms each pair is separated
by an interlayer coupling energy of order $\gamma=0.2\ eV$. (c):
multibarrier structure consisting of $N+2$
regions.}\label{Fig.system}
\end{figure}

We consider electron tunneling through a lateral one-dimensional
multibarrier in AA-stacked bilayer graphene as shown in Figure
\ref{Fig.system}(c), consisting of $N+2$ regions indexed by $r$.
The parameters for the regions represented by an odd index $r=2n_-+1$,
are the potential barrier {heights} $V_{2n_-+1}$,
the widths $d$ and the gaps $\Delta_{2n_-+1}$. However, for an
even index, we have $V_{2n_+}=\Delta_{2n_+}=0$, where
$n_\pm=\left\{0,\ 1,\
\cdots,\frac{N\pm1}{2}\right\}$, with the widths $a$.\\

\noindent 
The
charge carriers in the AA-stacked bilayer graphene are described, in each
region $r$, by the following four-band Hamiltonian \cite{Tabert}
\begin{equation}\label{eq1}
H_r=\left(%
\begin{array}{cccc}
  V_r+\Delta_r & v_F(p_x-ip_y) & \gamma & 0  \\
  v_F(p_x+ip_y) &  V_r-\Delta_r  & 0 & \gamma \\
  \gamma &  0 &  V_r+\Delta_r &  v_F(p_x-ip_y) \\
   0 & \gamma &  v_F(p_x+ip_y) &  V_r-\Delta_r  \\
\end{array}%
\right).
\end{equation}
The eigenstates of  $H_r$ are the four component spinors
$\Psi^{r}=\left(\Psi_{A}^{r}, \Psi_{B}^{r}, \Psi_{A^{'}}^{r},
\Psi_{B^{'}}^{r}\right)^T$, where $\Psi_{A(A^{'})}$ and
$\Psi_{B(B^{'})}$ are the envelope functions associated with the
probability amplitudes of the wave functions on the $A(A^{'})$
and $B(B^{'})$ sublattices of the {lower (upper)}
layer. In \eqref{eq1}, $p_{x,y}=-i\hbar \nabla_{x,y}$ is the
in-plane momentum relative to the Dirac point, $v_F = 10^6 m/s$ is
the Fermi velocity for electrons in single layer  graphene, $V_r$
is the potential heights and $\Delta_r$ is the gaps. In Figure
\ref{Fig.system}(a), we have the bilayer graphene consisting of
two layers of graphene having the structure AA-stacked, where the
A sublattice of the top layer is stacked directly above the same
sublattice of the bottom layer by the interlayer coupling
$\gamma=0.2\ eV$. 
 The unit cell is consisting of four
atoms labeled $A$, $B$ on the lower layer and $A^{'}$, $B^{'}$ on
the upper layer, see Figure \ref{Fig.system}(b).

We notice that our system is infinite along the
\textit{y}-direction since $[H_r,p_y]=0$ and then we can decompose
our spinor into
\begin{equation}\label{eq2}
\Psi^{r}(x,y)=e^{ik_yy}\psi^{r}(x,k_y).
\end{equation}
From the eigenvalue equation
$H_r\Psi^r=E\Psi^r$, we end up with 
 four linear differential
equations of the from
\begin{subequations}\label{eq3a}
\begin{align}{}
{-i\eta(\partial_x+k_y)\psi_{B}^{r}+\psi_{A^{'}}^{r}=(E-V_r-\Delta_r)\psi_{A}^{r}}\\
{-i\eta(\partial_x-k_y)\psi_{A}^{r}+ \psi_{B^{'}}^{r}=(E-V_r+\Delta_r)\psi_{B}^{r}}\\
{-i\eta(\partial_x+k_y)\psi_{B^{'}}^{r}+ \psi_{A}^{r}=(E-V_r-\Delta_r)\psi_{A^{'}}^{r}}\\
{-i\eta(\partial_x-k_y)\psi_{A^{'}}^{r}+\psi_{B}^{r}=(E-V_r+\Delta_r)\psi_{B^{'}}^{r}}
\end{align}
\end{subequations}
where we have introduced the length scale $\eta=\frac{\hbar
v_F}{\gamma}\thickapprox 3.29~nm$. From now on we switch to
dimensionless quantities by measuring all energy terms in units
of the interlayer coupling $\gamma$, namely $E \longrightarrow
\frac{E}{\gamma}$, $V_r \longrightarrow \frac{V_r}{\gamma}$ and
$\Delta_r\longrightarrow \frac{\Delta_r}{\gamma}$. We combine the
above equations to eliminate the unknown ones at time to end up with
 the following differential equation
\begin{equation}
\left[\partial_{x}^2+(k_{r}^{\tau})^2\right]\psi_{B}^{r}(x,k_y)=0
\end{equation}
where the wave vector along the \textit{x}-direction reads as
\beq
k_{r}^{\tau}=\sqrt{-k_{y}^2+\eta^{-2}\left(\left(E-V_r-\tau\right)^2-\Delta_{r}^2\right)}
\eeq
 and $\tau$ is
the cone index, with $\tau=-1$ ($\tau=+1$) for the lower (upper)
cone. Then
we have $V_{2n_-+1}$ and $\Delta_{2n_-+1}$ equal to $V_1$ and
$\Delta_1$, respectively. For the regions where  $V_1 =
\Delta_1=0$, we find the wave vector
\beq
k_{0}^{\tau}=\sqrt{-k_{y}^2+\eta^{-2}\left(E-\tau\right)^2}
\eeq
and the
corresponding energy band  
is 
\begin{equation}\label{energybands0}
E^{s_0,\tau}=
\tau+s_0\sqrt{\eta^2\left((k_{0}^{\tau})^2+(k_{y})^2\right)}.
\end{equation}
However generally, for any region we can deduce the energy from
previous analysis as
\begin{equation}\label{energybands}
E^{s,\tau}=V_1+
\tau+s_{1}\sqrt{\eta^2\left((k_{1}^{\tau})^2+(k_{y})^2\right)+\Delta_{1}^2}
\end{equation}
where $s_{1}$ is the chirality index of a quasiparticle in the
barrier regions ($V_r=V_{1}$, $\Delta_r=\Delta_{1}$) and in
outside ($V_r=\Delta_r=0$) is $s_0$, with $s_{1}=+1$ (or $s_0
=+1$) and $s_{1}=-1$ (or $s_0 =-1$) are the electron-like and
hole-like band indexes. It is important to note that in
\eqref{energybands}, for $\gamma\longrightarrow 0$, the energy bands
will be reduced to
\beq
E^{s}=V_1+
s_1\sqrt{\hbar v_F((k_{1}^{\tau})^2+(k_{y})^2)+\Delta_{1}^2}
\eeq
which is a degenerate energy and {corresponding to two monolayers.
It is clear that this degeneracy can be lifted by taking into account
the coupling between
the two monolayers, i.e. $\gamma \neq 0$ \cite{Anna}.}\\

\begin{figure}[H]
 \centering
 \includegraphics[width=5cm, height=4cm]{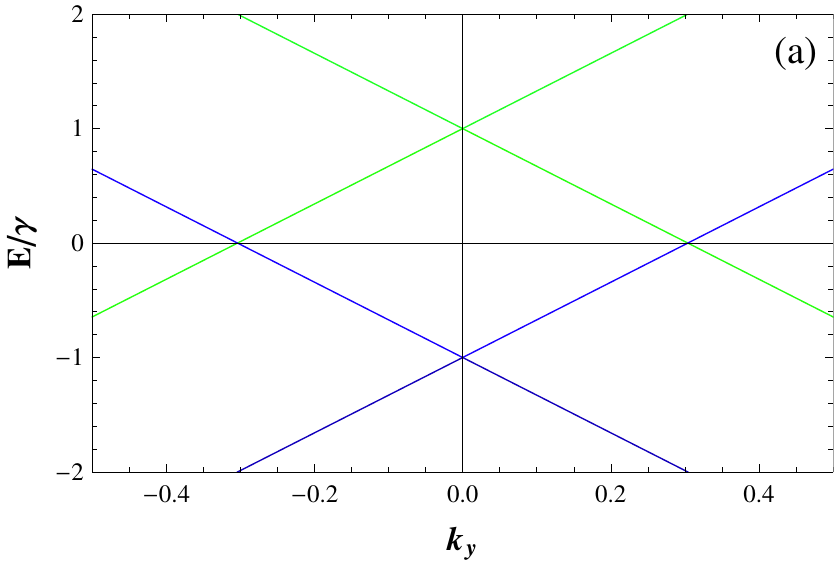}
~~\includegraphics[width=5cm, height=4cm]{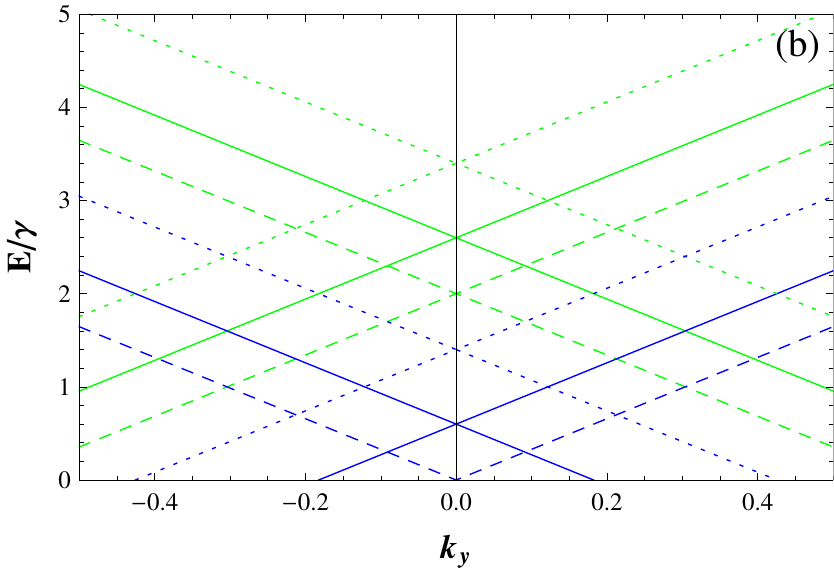}
~~\includegraphics[width=5cm, height=4cm]{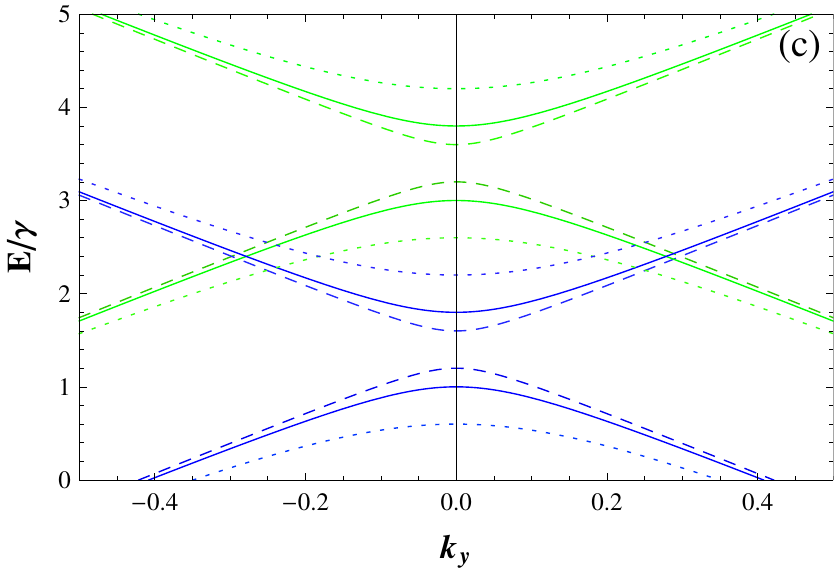}
 \caption{\sf The energy bands as a function of the momentum $k_y$, where
the blue line (green line) corresponds to the lower cone (upper
cone). (a): for $V_{1}=\Delta_{1}=0$. (b): for $\Delta_{1}=0$ and
$V_{1}=(\gamma,\ 1.6 \gamma,\ 2.4 \gamma)$ correspond to (dashed
line, solid line, dotted line). (c): for $V_{1}=2.4\gamma$ and
$\Delta_{1}=(0.2 \gamma,\ 0.4 \gamma,\ 0.8 \gamma)$ correspond to
(dashed line,
 solid line, dotted line).} \label{Fig.energy}
\end{figure}

In Figure \ref{Fig.energy}, we plot the energy bands as a function
of the momentum $k_y$ for the lower cone {$\tau=-1$
(blue line) and the upper cone $\tau=1$ (green line).}
In Figures \ref{Fig.energy}(a) and \ref{Fig.energy}(b) the energy
bands are plotted for $\Delta_{1}=0$ while in Figure
\ref{Fig.energy}(c) for $\Delta_{1} \neq 0$. {It is
evident that in Figure \ref{Fig.energy} the energy bands are just
the double copies of monolayer graphene bands, with one Dirac
point is shifted to wards positive  ($\tau=+1$) and the other
towards negative ($\tau=-1$) energies.
For zero gap the spectrum is linear (Figures \ref{Fig.energy}(a)
and \ref{Fig.energy}(b)), however for a finite gap the spectrum is
parabolic (Figures \ref{Fig.energy}(c)).}
In
Figure \ref{Fig.energy}(b), we plot the energy bands for three
different values of the potential heights $V_{1}=(\gamma,\
1.6\gamma,\ 2.4\gamma)$. From this, we can observe that when we
increase the potential heights $V_{1}$, the energy bands increase
upwards.
To see the effect of the gap, we plot the energy bands for three
different values of $\Delta_{1}=(0.2\gamma,\ 0.4\gamma,\
0.8\gamma)$ and for $V_{1}=2.4\gamma$, the results are shown in
Figure \ref{Fig.energy}(c). {It is clearly seen that a gap appears
at each of the two Dirac points {($V_{1}+\tau-\Delta_{1} < E <
V_{1}+\tau+\Delta_{1}$)}. When we increases $\Delta_{1}$, the
width of these gaps increases as well.


Figure \ref{Fig.regionenergy} represents the band structure of
AA-stacked bilayer graphene for zero and finite gap. For $r=0$,
which corresponds to the incident region, the electron states can be
subdivided into two regimes. The first one is where both $\tau=\pm
1$ electrons are electron-like, the second one is where $\tau=+1$
electrons are hole-like while $\tau=-1$ electrons are
electron-like. However, for the transmission region ($r=N + 1$),
we have two regimes as we shall show in the incident region. Note
 that we have four transmissions that are related to the
band structures on the incident and transmission region. In
addition, we notice that for the intracone transitions
(\textit{i.e.} $\tau \longrightarrow \tau$ processes), there exist
four cases for transitions across the barrier:
\begin{itemize}
 \item  Electron in
regime I $\longrightarrow$ Electron in regime I\ \ \ \ (1)
\item  Electron in
regime I $\longrightarrow$ Electron in regime II\ \ \ (2)
\item  Electron in
regime II $\longrightarrow$ Electron in regime I \ \ \ (3)
\item  Electron in
regime II $\longrightarrow$ Electron in regime II\ \ \ (4)
\end{itemize}
However, all
intercone transitions (\textit{i.e.} $\tau \rightarrow -\tau$
processes) are strictly forbidden due to the orthogonality of
electron wave functions with a different cone index
\cite{Sanderson}. As already mentioned above, the transitions
depend on the incident and transmission regions, so they depend
anywhere in the presence or absence of the gap. For each of the
four cases for transitions, the transmission can be calculated
using the same method.

\begin{figure}[h!]
 \centering
 \includegraphics[width=14cm, height=2.5cm]{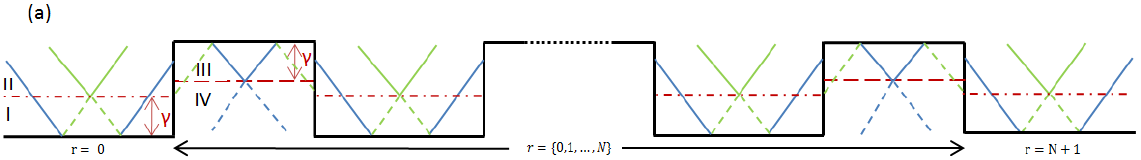}
\\
 \includegraphics[width=14cm, height=2.5cm]{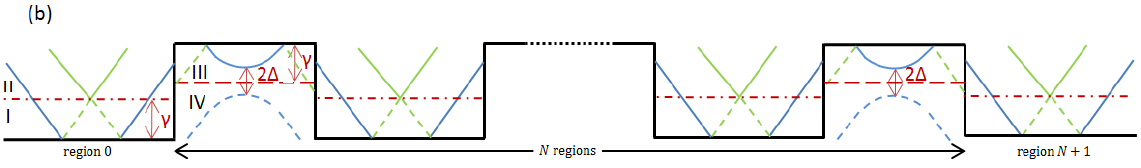}
 \caption{\sf Schematic representation of the band structures. (a): for the barriers structure with zero gap.
 (b): for the barriers structure with finite gap.}\label{Fig.regionenergy}
\end{figure}

\section{Transmission probability}

Next we shall show in detail the calculation of the transmission
probability of electrons across the periodic barrier structure in
our AA-stacked bilayer graphene system. The wave function solution
of the system in each region $r$ can then be written in terms of a
matrix form as
\begin{equation}
\psi^r=G_r \cdot M_r\cdot A_r
\end{equation}
where different matrices are given by
\begin{equation}
G_{r}=\left(%
\begin{array}{cc}
s_r\tau f_{r}^{\tau,+} &  s_r\tau f_{r}^{\tau,-}\\ \tau  &  \tau
\end{array}%
\right),\qquad M_{r}=\left(%
\begin{array}{cc}
  e^{ik_{r}^{\tau}x} & 0  \\
  0 &  e^{-ik_{r}^{\tau}x} \\
\end{array}%
\right),\qquad A_{r}^\tau=\left(%
\begin{array}{cc}
  \alpha_{r}^\tau \\
  \beta_{r}^\tau  \\
  \end{array}%
\right)
\end{equation}
such that $f_{r}^{\tau,\pm}$ reads as \beq f_{r}^{\tau,\pm}=\pm
\sqrt{\frac{E-V_1-\tau+\Delta_1}{E-V_1-\tau-\Delta_1}}e^{\mp
i\phi_{r}^\tau} \eeq
 and the phase is defined by
 \beq
\phi_{r}^\tau=\arctan(k_y/k_{r}^\tau).
\eeq
We are interested
in the normalization coefficients, namely the components of $A_{r}^\tau$,
on the both sides of the multibarrier structure. In other words,
for the incident region ($r = 0$) and  transmission region ($r = N+1$ ), respectively, we have

\beq
A_{0}^\tau =\left(%
1,r^\tau \right)^T, \qquad A_{N+1}^\tau =\left(%
t^\tau ,0\right)^T
\eeq
where $r^\tau $ and $t^\tau $ are the
reflection and transmission coefficients of each cone ($\tau=\pm
1$), respectively.
We use two different matrix notation written as
\begin{equation}
G_{2n_+}\cdot M_{2n_+}=G_{0}\cdot M_{0}, \qquad
G_{2n_-+1}\cdot M_{2n_-+1}=G_{1}\cdot M_{1}
\end{equation}
where we have $V_1=\Delta_1=0$ in $r=2n_+$ and $V_1\neq 0$ and
$\Delta_1\neq 0$ for $r=2n_-+1$, with $n_\pm=\left\{0,\ 1,\
\cdots,\frac{N\pm 1}{2}\right\}$. Using the boundary conditions and the
transfer matrix method, we can connect $A_{0}^{\tau}$ with
$A_{N+1}^{\tau}$ through the matrix $\zeta$ 
\begin{equation}
\zeta=\prod_{j=0}^{n}M_{0}^{-1}\left[j(d+a)\right]\cdot
G_{0}^{-1}\cdot G_{1}\cdot M_{1}\left[j(d+a)\right]\cdot
M_{1}^{-1}\left[(j+1)d+ja\right]\cdot G_{1}^{-1}\cdot G_{0}\cdot
M_{0}\left[(j+1)d+ja\right].
\end{equation}
Consequently, we have two channels for the transmission probability in each individual cone, which are given by
\beq
T^\tau=\frac{1}{[\zeta_{11}]^2}
\eeq
where $\zeta_{11}$ is a  element of the matrix $\zeta$.\\ 


\begin{figure}[h!]
 \centering
  \includegraphics[width=6.5cm, height=4.5cm]{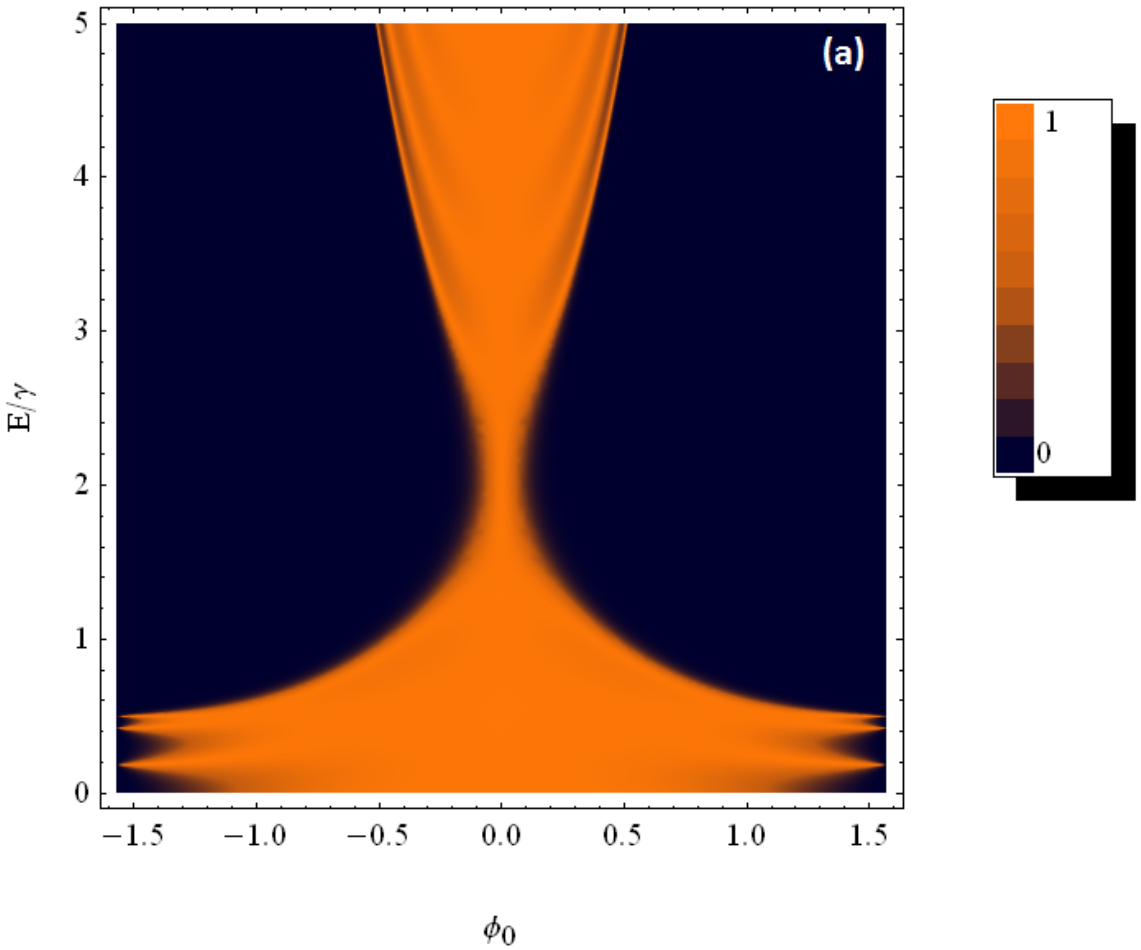}
 ~~~~~~ \includegraphics[width=6.5cm, height=4.5cm]{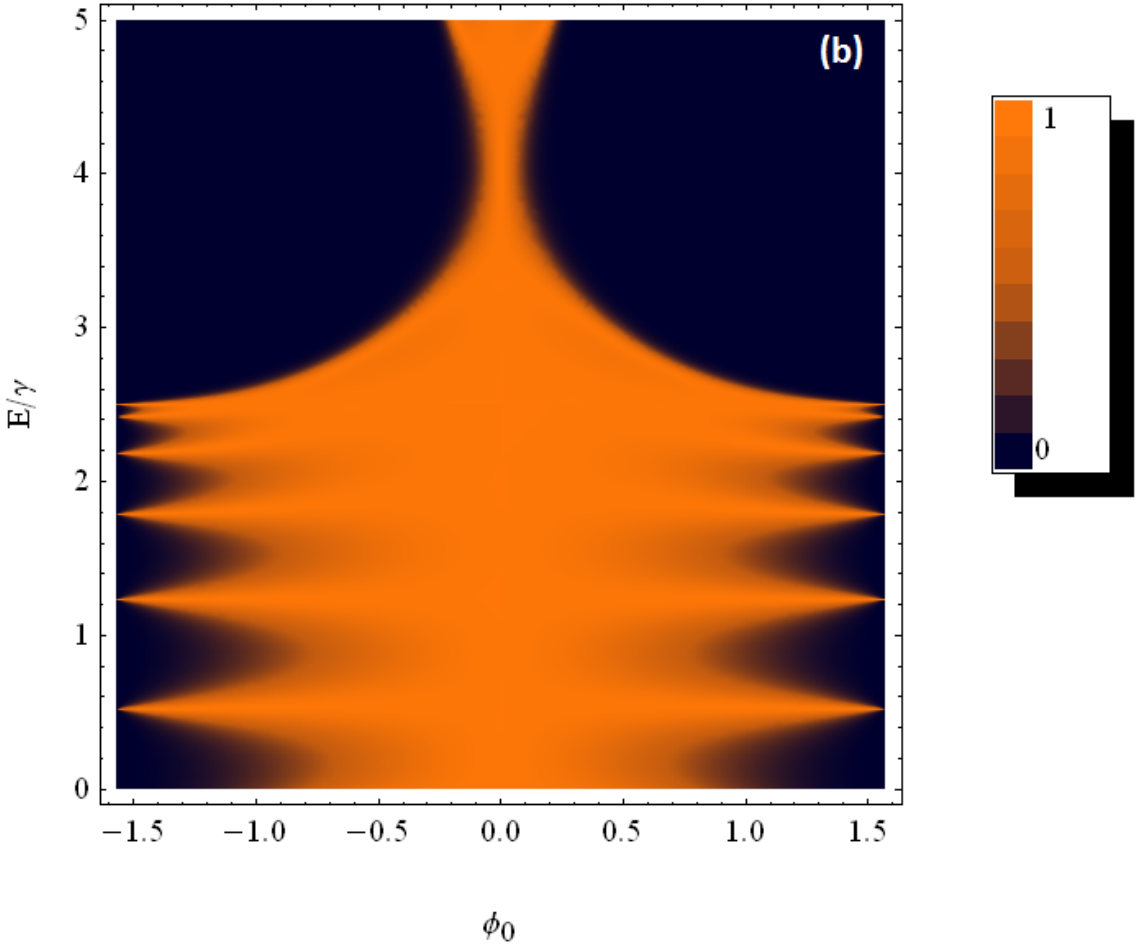}
 \\ \includegraphics[width=6.5cm, height=4.5cm]{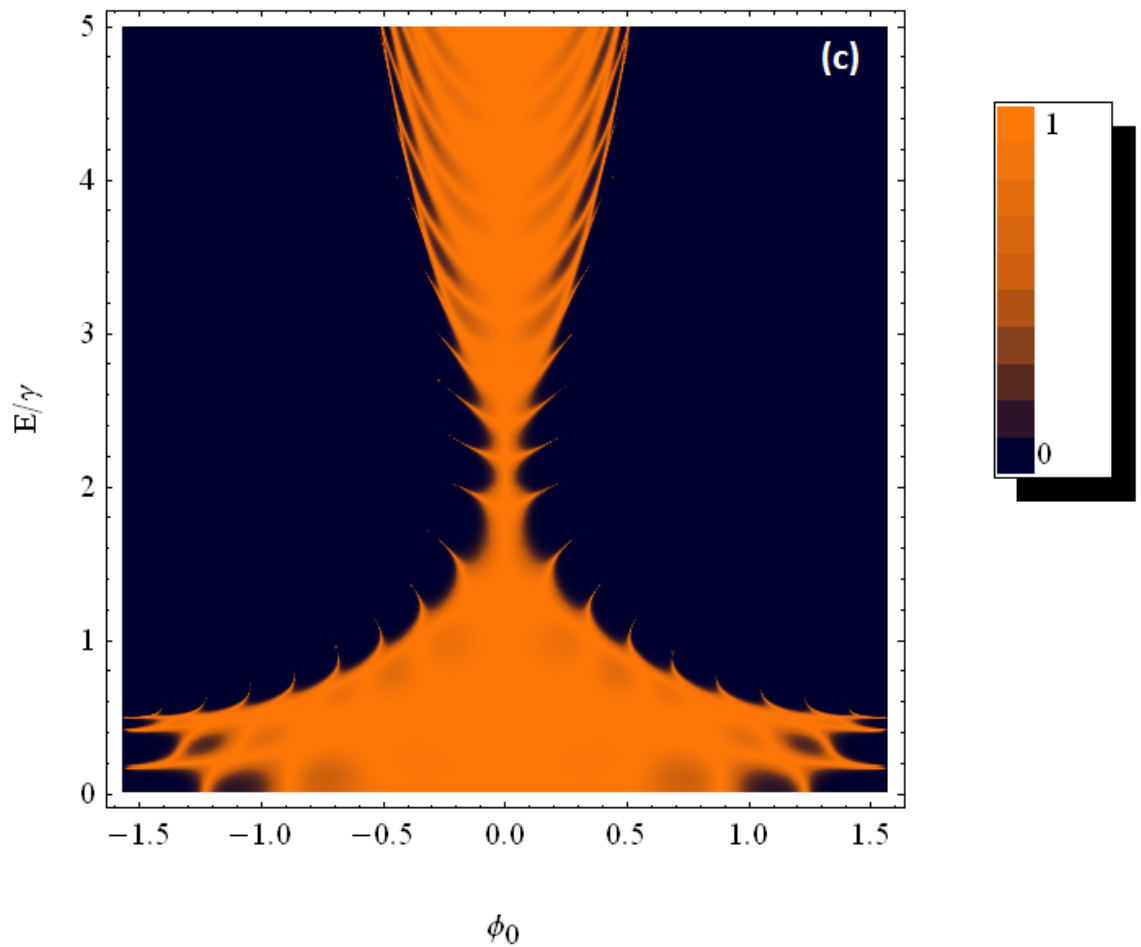}
 ~~~~~~ \includegraphics[width=6.5cm, height=4.5cm]{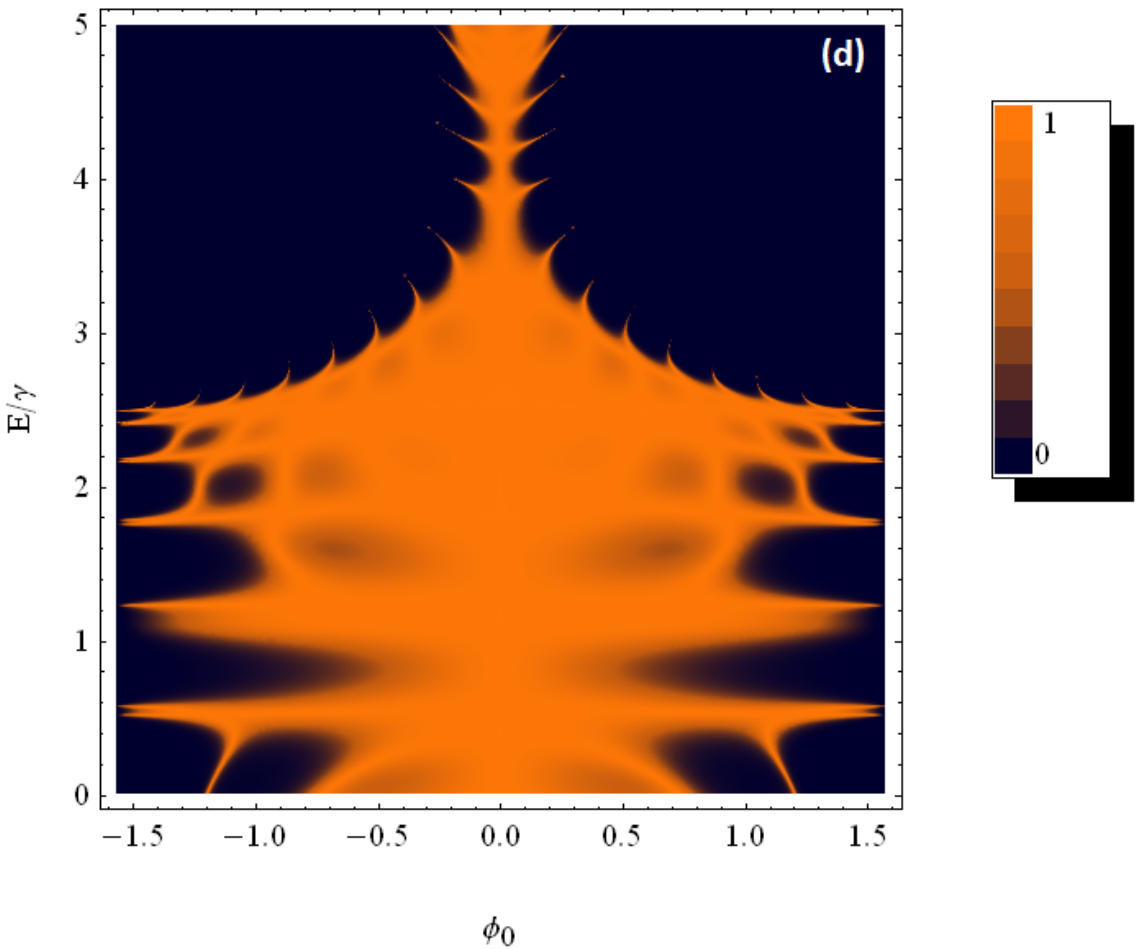}
 \caption{\sf Density plot of transmission probability as a function of
the incident angle $\phi_0$ and its energy $E$, for $V_1=3\gamma$,
$a=2d=30~nm$ and $\Delta_1=0\gamma$. (a)/(b) for single barrier
with $\tau=-1$/$\tau=1$. (c)/(d) for double barrier with
$\tau=-1$/$\tau=1$. } \label{fig.FB2}
\end{figure}

To allow for a suitable interpretation of our main results, we
compute numerically the transmission probabilities for $\tau=1$
and $\tau=-1$. In what follow we start by studying the
transmission probability for single, double barrier and finite
periodic barrier {structures}. We first consider the case of
single and double barrier geometry with zero gap. In Figure
\ref{fig.FB2} we show the density plot of the transmission
probability as a function of the incident angle and its energy.
The different colors from blue to orange correspond to different
values of the transmission from 0 to 1. It is important to note
that in the case of AA-stacked bilayer graphene the band structure
is composed of two Dirac cones shifted by $\tau=-1$ (Figures
\ref{fig.FB2}(a) and \ref{fig.FB2}(c)) and $\tau=1$ (Figures
\ref{fig.FB2}(b) and \ref{fig.FB2}(d)). In addition, the
transmission probabilities, for $\tau=-1$ and $\tau=1$, have the
same form as that in the case of single layer  graphene
\cite{Azarova}. In both cases of single layer  and AA-stacked
bilayer graphene, around the point $\phi_0=0$, we have perfect
transmission with a manifestation of the Klein tunneling effect
\cite{Klein}. Until now we notice that there is some similarity
between single layer  and AA-stacked bilayer graphene. However,
for AA-stacked bilayer graphene both the electrons and holes have
the same chirality index  $s=\pm1$ while in the case of single
layer  graphene the electron has always $+1$ and $-1$ for the
hole. For the transmission of the lower layer (Figures
\ref{fig.FB2}(a) and \ref{fig.FB2}(c)) the Dirac points correspond
to $E=V_1-1$ and for the upper layer (Figures \ref{fig.FB2}(b) and
\ref{fig.FB2}(d)) correspond to $E=V_1+1$. The results for a
symmetrical double barrier structure of width $d$ and interbarrier
separation $a$, are shown in Figures \ref{fig.FB2}(c) and
\ref{fig.FB2}(d). Compared to the results seen above for single
barrier, we observe the appearance of
peaks in transmission probability.\\

\begin{figure}[h!]
 \centering
  \includegraphics[width=6.5cm, height=4.5cm]{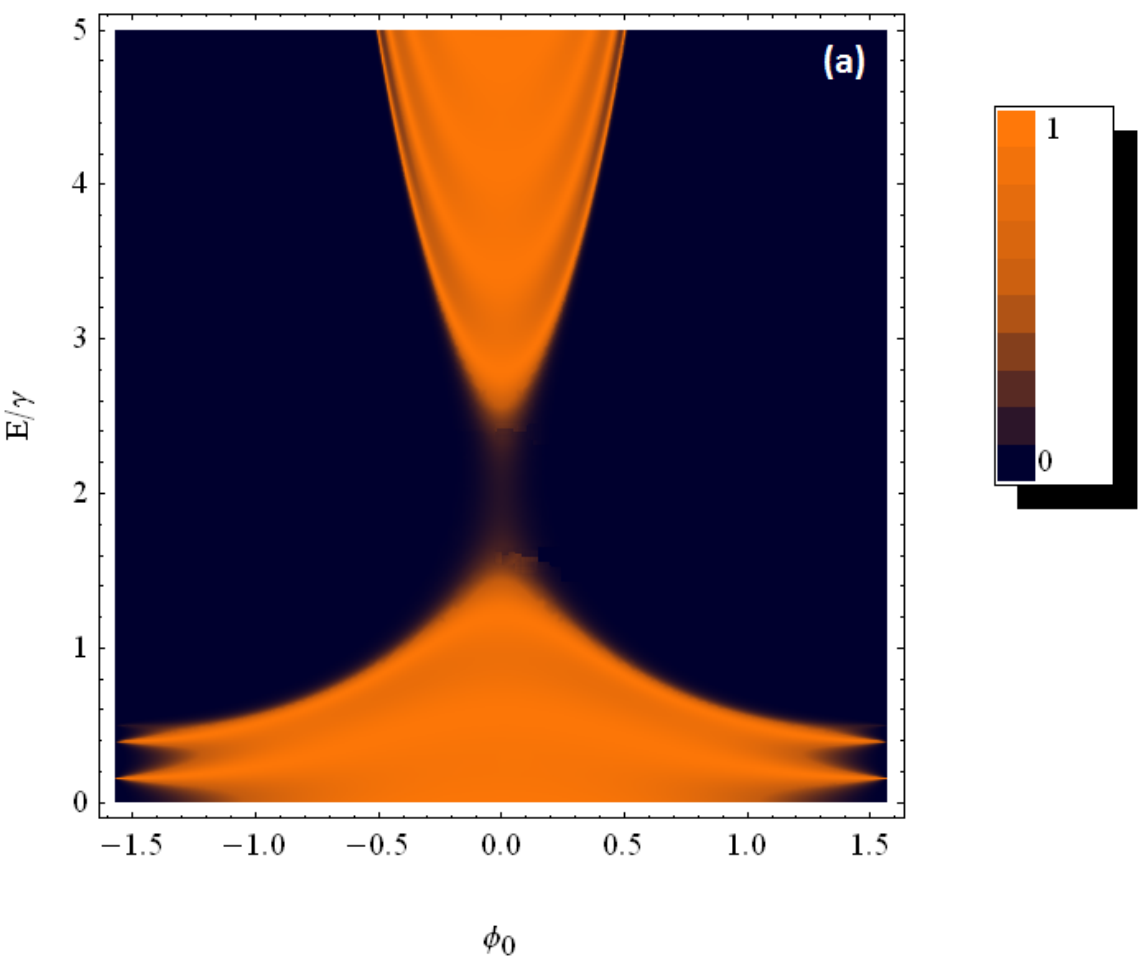}
 ~~~~~~ \includegraphics[width=6.5cm, height=4.5cm]{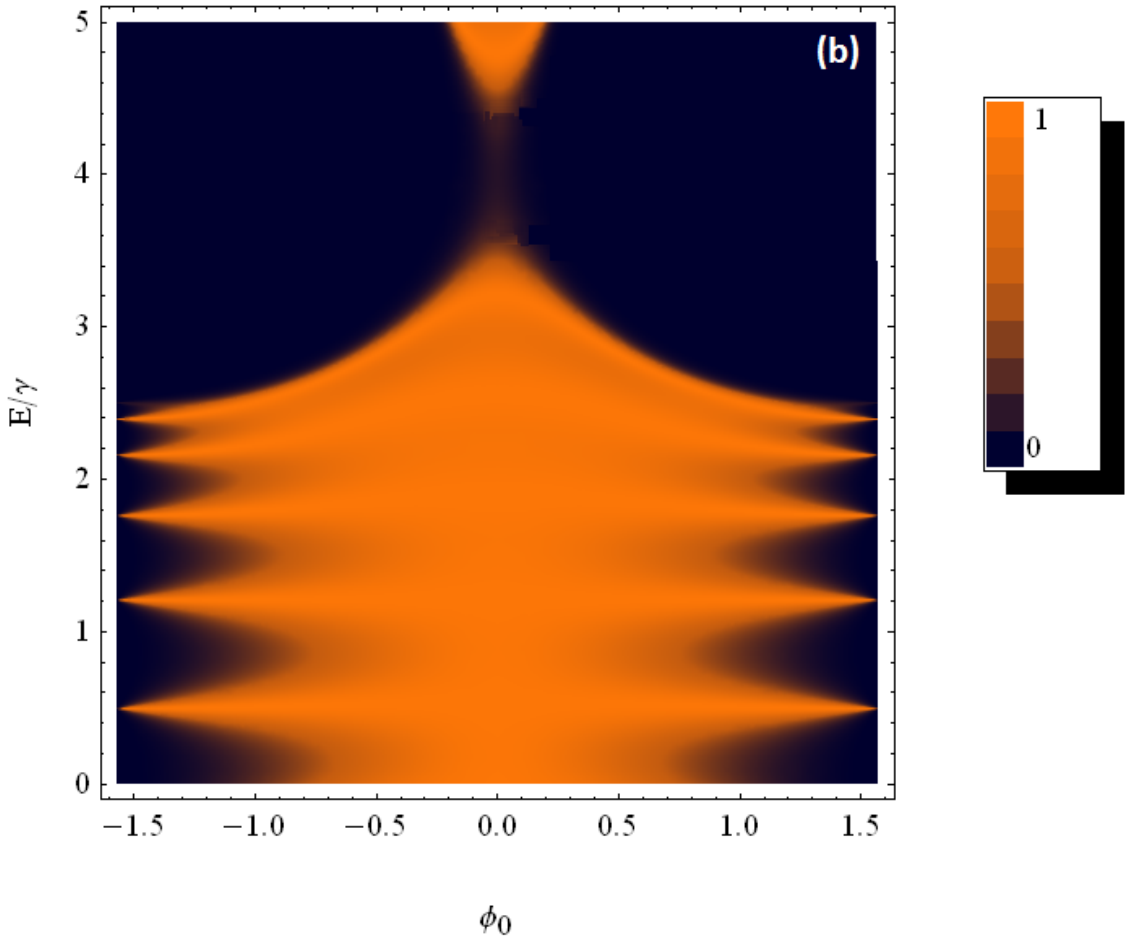}
 \\ \includegraphics[width=6.5cm, height=4.5cm]{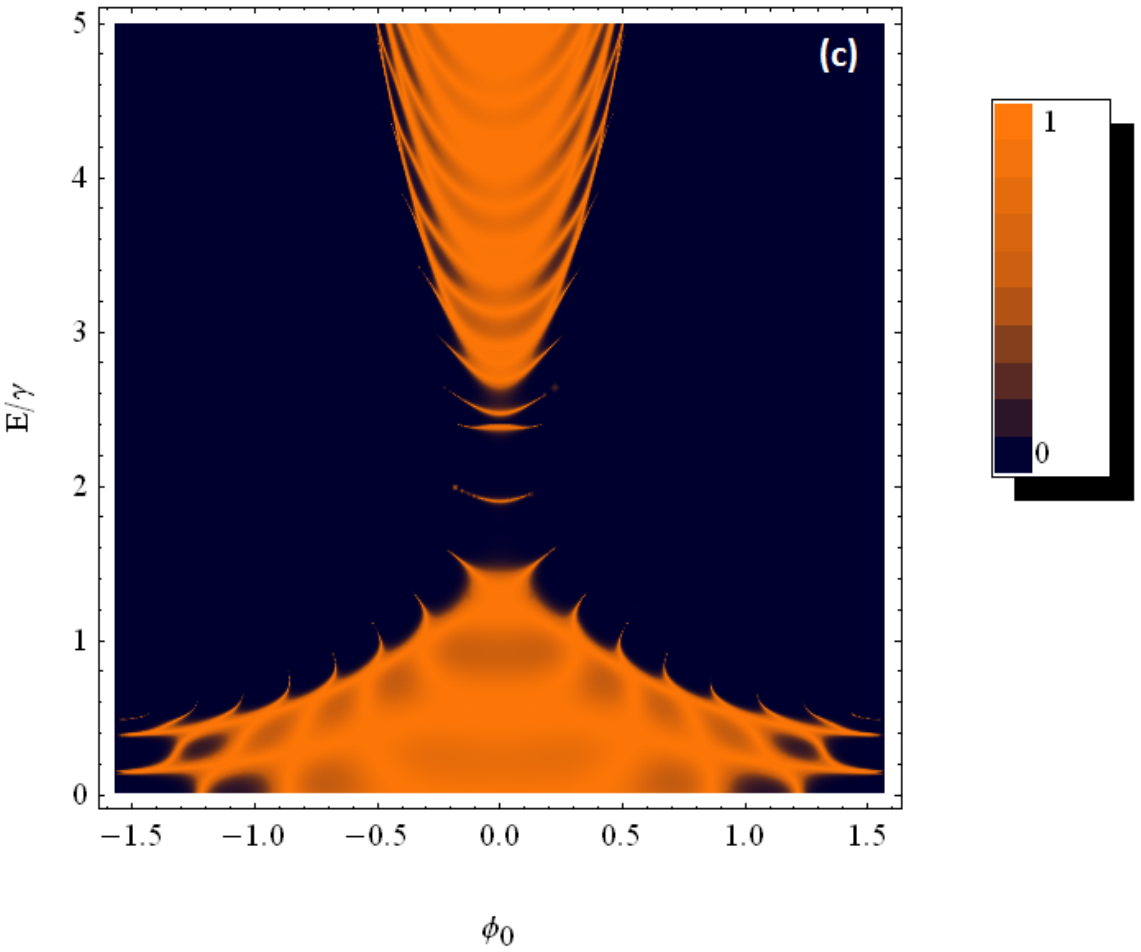}
 ~~~~~~ \includegraphics[width=6.5cm, height=4.5cm]{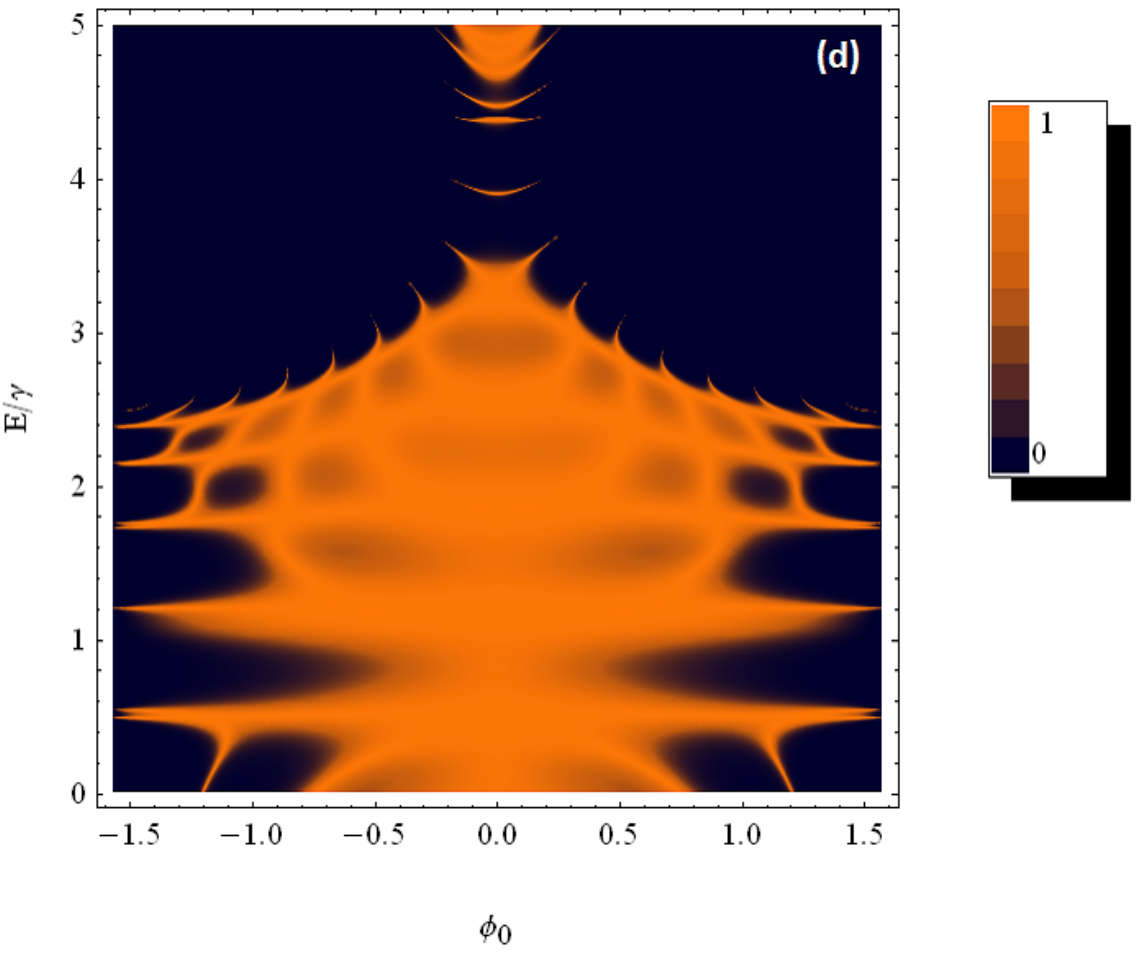}
 \caption{\sf Density plot of transmission probability as a function of
the {incident angle} $\phi_0$ and its energy $E$, for
$V_1=3\gamma$, $a=2d=30~nm$ and $\Delta_1=0.4\gamma$. (a)/(b) for
single barrier with $\tau=-1$/$\tau=1$. (c)/(d) for double barrier
with $\tau=-1$/$\tau=1$.} \label{fig2.FB2}
\end{figure}

Now let us see what will happen if we introduce a gap in the band
structures. To do this, we extend the results presented in Figure
\ref{fig.FB2} to the case $\Delta_1=0.4 \gamma$ to get Figure
\ref{fig2.FB2}. It is clearly shown that, in contrast to the
single barrier case (Figures \ref{fig2.FB2}(a) and
\ref{fig2.FB2}(b)), there are full transmission resonances, in the
case of double barrier (Figures \ref{fig2.FB2}(c) and
\ref{fig2.FB2}(d)), inside the gap {($V_1+\tau -
\Delta_1\ < E <\ V_1+\tau + \Delta_1$)}. Note that, the
transmission resonances are resulting from the available states in
the well between the barriers. As already mentioned above, for a
gapless graphene we have perfect transmission that is a
manifestation of Klein tunneling effect. In addition, it may be
noted that the opening gap in
the barrier region suppresses this effect.\\

\begin{figure}[H]
 \centering
  \includegraphics[width=6.5cm, height=4.5cm]{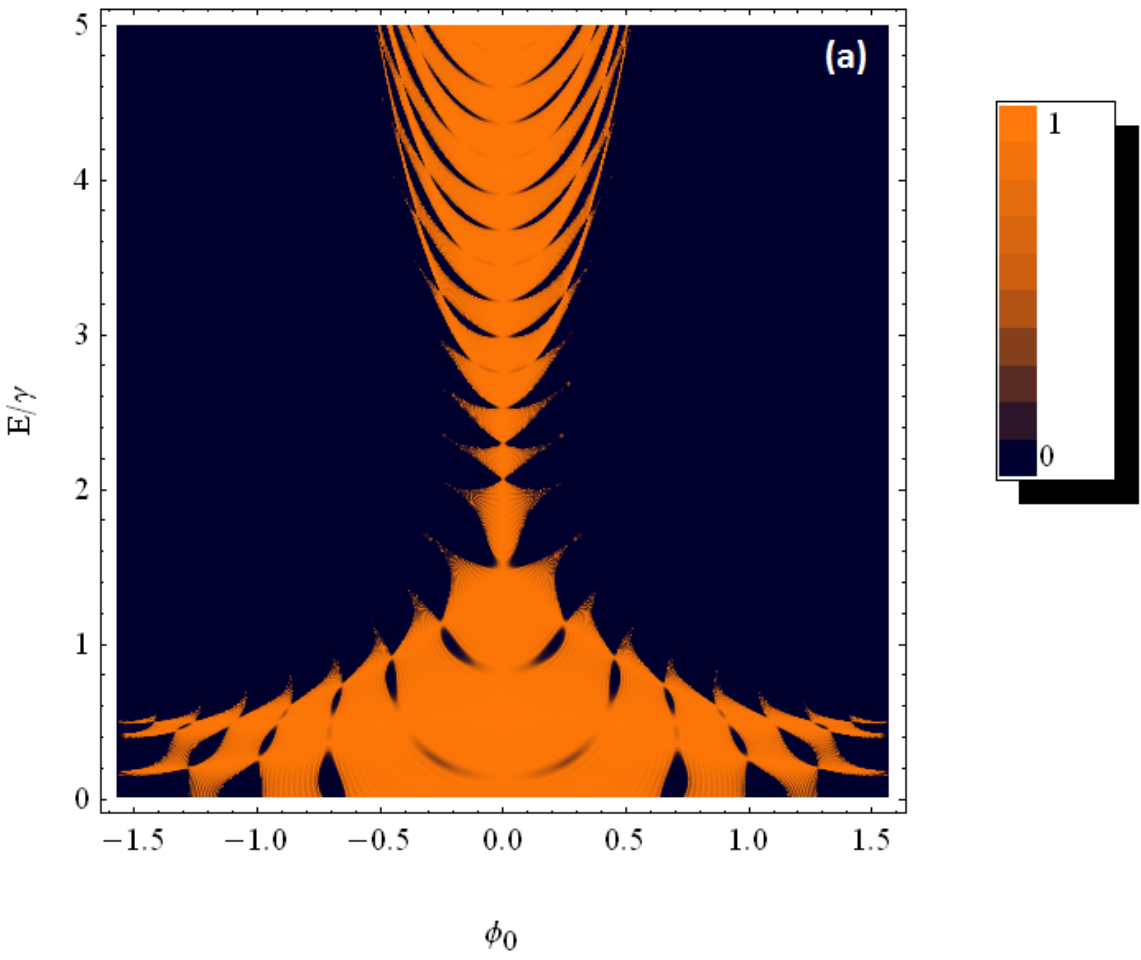}
 ~~~~~~ \includegraphics[width=6.5cm, height=4.5cm]{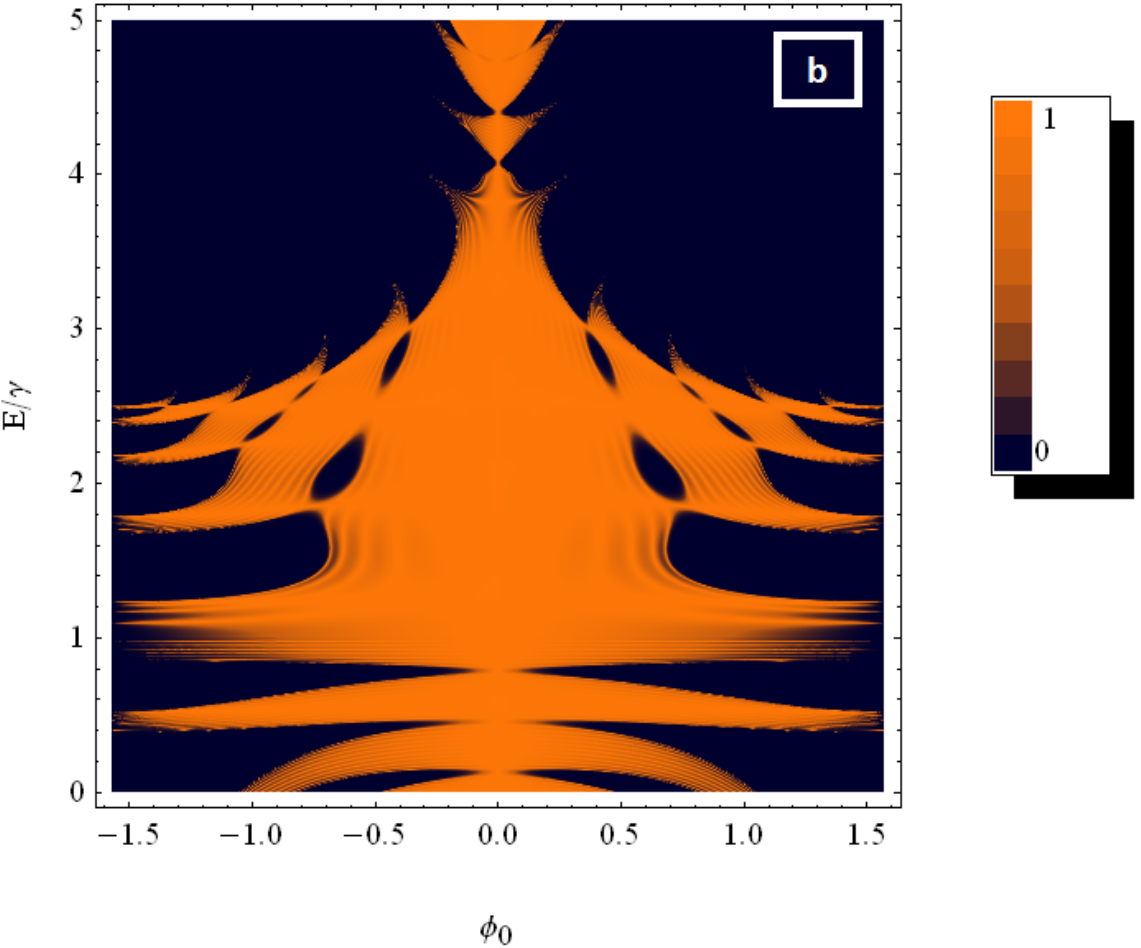}
 \\ \includegraphics[width=6.5cm, height=4.5cm]{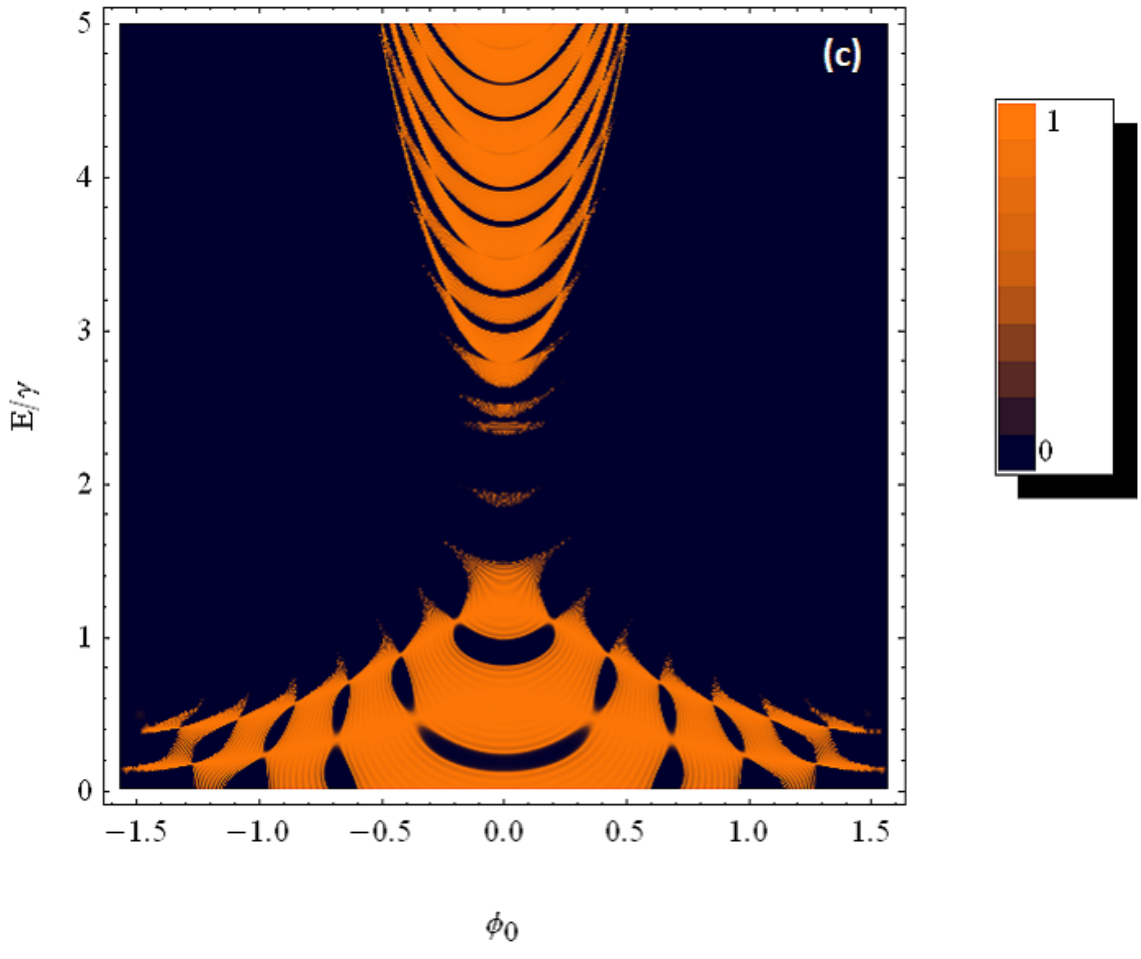}
 ~~~~~~ \includegraphics[width=6.5cm, height=4.5cm]{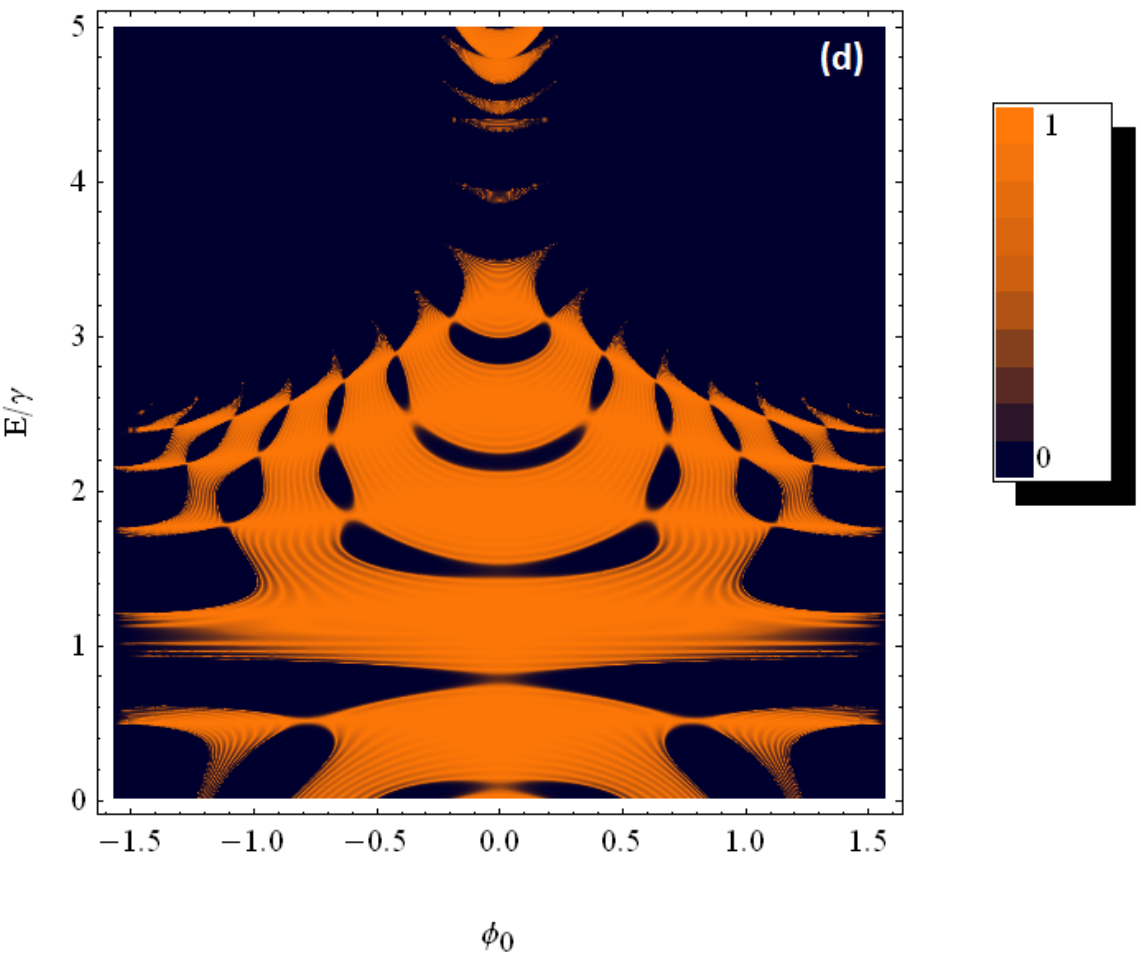}
 \caption{\sf Density plot of transmission probability as a function of
the {incident angle} $\phi_0$ and its energy $E$, for finite
periodic barrier structure that contain $25$ regions, with $V_1=3$
and $a=2 d=30\ nm$. (a)/(c) for  $\tau=-1$ with $\Delta_1=0$
/{$\Delta_1=0.4 \gamma$}. (b)/(d) for $\tau=1$ with
$\Delta_1=0$/{$\Delta_1=0.4 \gamma$.}} \label{fig3.FB2}
\end{figure}

Next we consider the finite periodic barrier structure that
contain $25$ regions. In Figure \ref{fig3.FB2} we show the
transmission probabilities, associated with $\tau=\pm 1$, as a
function of the {incident angle} $\phi_0$ and its energy $E$, for
$V_1=3$ and $a = 2 d = 30\ nm$. Figures \ref{fig3.FB2} (a)/(c) for
$\Delta_1=0$/{$\Delta_1=0.4 \gamma$} and $\tau= - 1$ while Figures
\ref{fig3.FB2} (b)/(d) for $\Delta_1=0$/{$\Delta_1=0.4 \gamma$}
and $\tau= + 1$. It is evident that the band structure of such
barrier can have more than one Dirac points located at $E=V_1-1$
for the lower cone (Figure \ref{fig3.FB2}(a)) and at $E=V_1+1$ for
the upper cone (Figure \ref{fig3.FB2}(b)). It is important to note
that the location of Dirac point does not depend on the number of
barriers. In fact, the position of the Dirac point is the same for
single, double and the periodic barrier structures. However,
increasing the number of barrier leads to the occurrence of
resonance peaks in the transmission probability. These resonance
peaks correspond to the positions of new cone index like
associated with $\tau=\pm 1$ (Figures \ref{fig3.FB2}(a) and
\ref{fig3.FB2}(b)) in the finite periodic barrier structures. We
should emphasize that for $\tau\longrightarrow 0$, our system will
be reduced to the case of single layer graphene
\cite{Maksimova,Azarova}. From Figure \ref{fig3.FB2}(c) and
\ref{fig3.FB2}(d), we can clearly see that the presence of the gap
leads to suppression of the Klein tunneling effect \cite{Klein}.
In addition, it is important to note that the number of peaks
presented inside the transmission gap {($V_1+\tau - \Delta_1\ < E
<\ V_1+\tau + \Delta_1$)}, in the case of double barrier structure
(Figure \ref{fig2.FB2}(c) and \ref{fig2.FB2}(d)), is the same as
that corresponding to the transmission trough finite periodic
barrier structure.

\section{Conductance}

{Basing on the obtained results above regarding the
transmission probabilities of our system, we will see how 
the two conductance ($G^\tau$) of each cone channel
($\tau=\pm 1$) will behave.
For this purpose,} we evaluate $G^\tau$ by using the
{Landauer-B\"{u}ttiker} \cite{Blanter} formula
\begin{equation}\label{condtau}
G^\tau=G_0\int_{0}^{\frac{\pi}{2}}T^\tau\left(E,\phi_0\right)\cos\phi_0\
d\phi_0
\end{equation}
where the unit conductance
\beq
G_0=N L_y k_F e^2/h\pi
\eeq
with the factor $N=4$ is due to the
spin and valley degeneracy,  $L_y$ is the width of
the sample in the $y$-direction and
\beq
k_F=\sqrt{k_{y}^2+(k_{0}^\tau)^2}=s_0\eta^{-1}(E-\tau).
\eeq
The
total conductance  $G_t$ {is defined as the sum of
the conductance channels in each individual cone ($G^\tau$)}
such as \cite{Sanderson}
\begin{equation}\label{condtotal}
G_t=\frac{1}{2}\left(G^+ +G^-\right)
\end{equation}
and the factor $1/2$ is required because the total conductance is
contribution of the two cones. {Whereas, in the case of
AB-stacked bilayer graphene, there are four transmissions
channels, then the total conductance is the sum of four
conductances channels \cite{Duppen}.}
In the forthcoming analysis, {we evaluate
numerically the conductance in AA-stacked} bilayer graphene.\\

\begin{figure}[h]
 \centering
  \includegraphics[width=6.5cm, height=4.5cm]{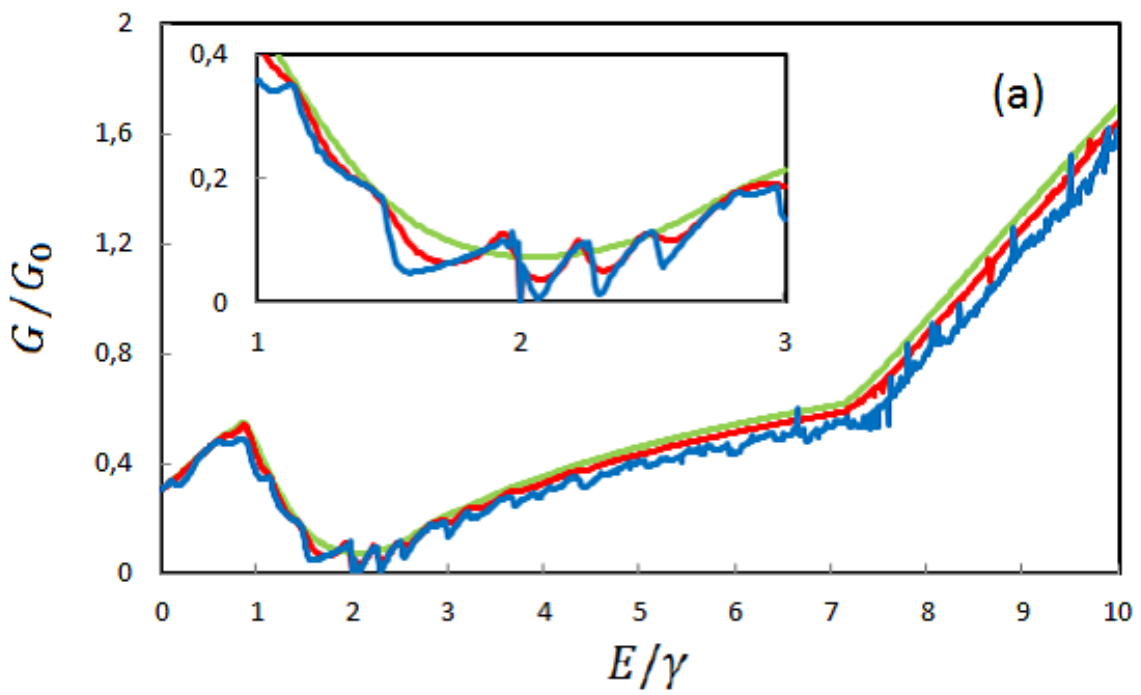}
 ~~~~~~\includegraphics[width=6.5cm, height=4.5cm]{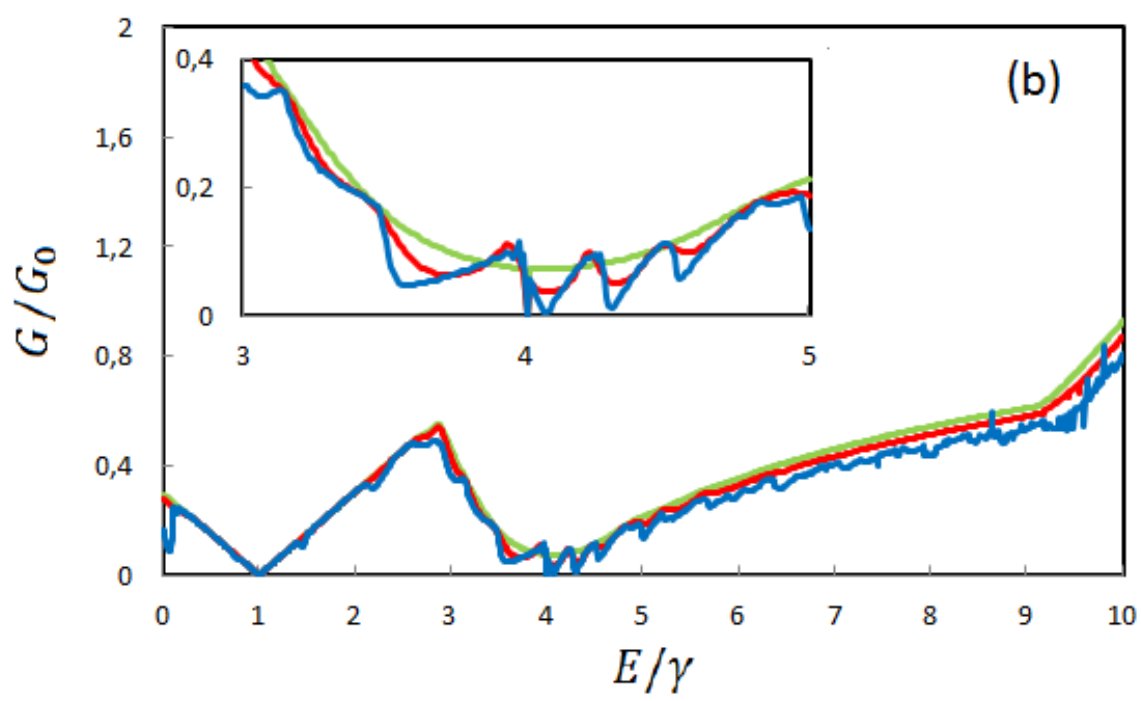}
 \\\includegraphics[width=6.5cm, height=4.5cm]{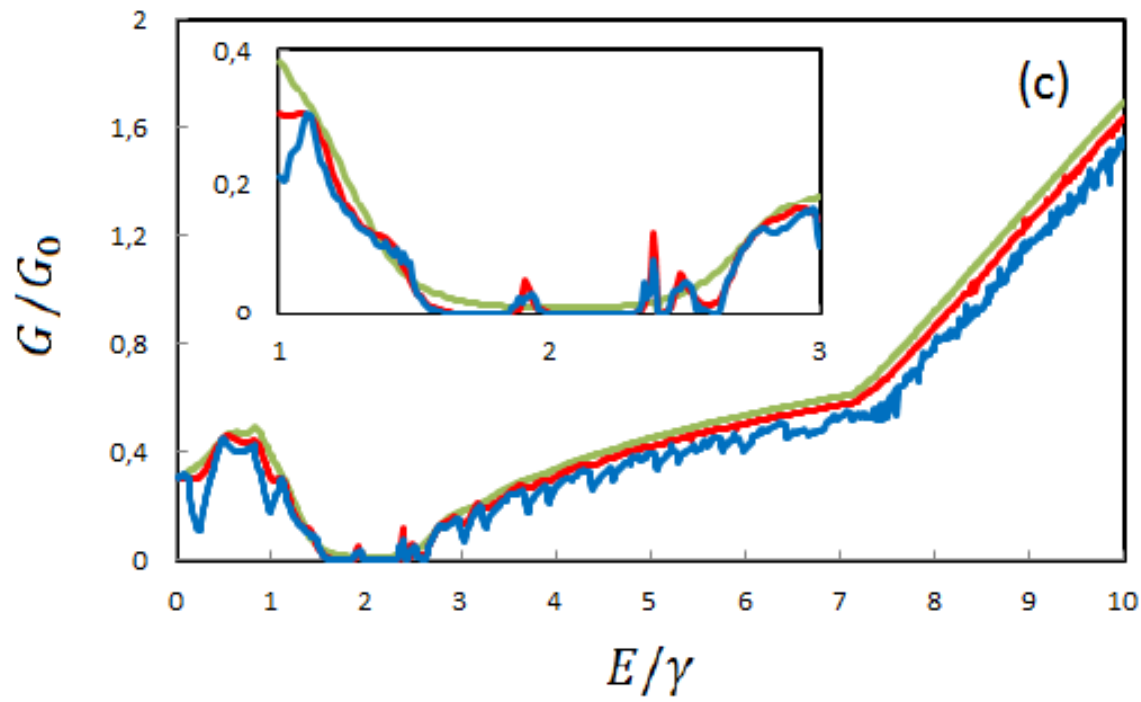}
~~~~~~\includegraphics[width=6.5cm, height=4.5cm]{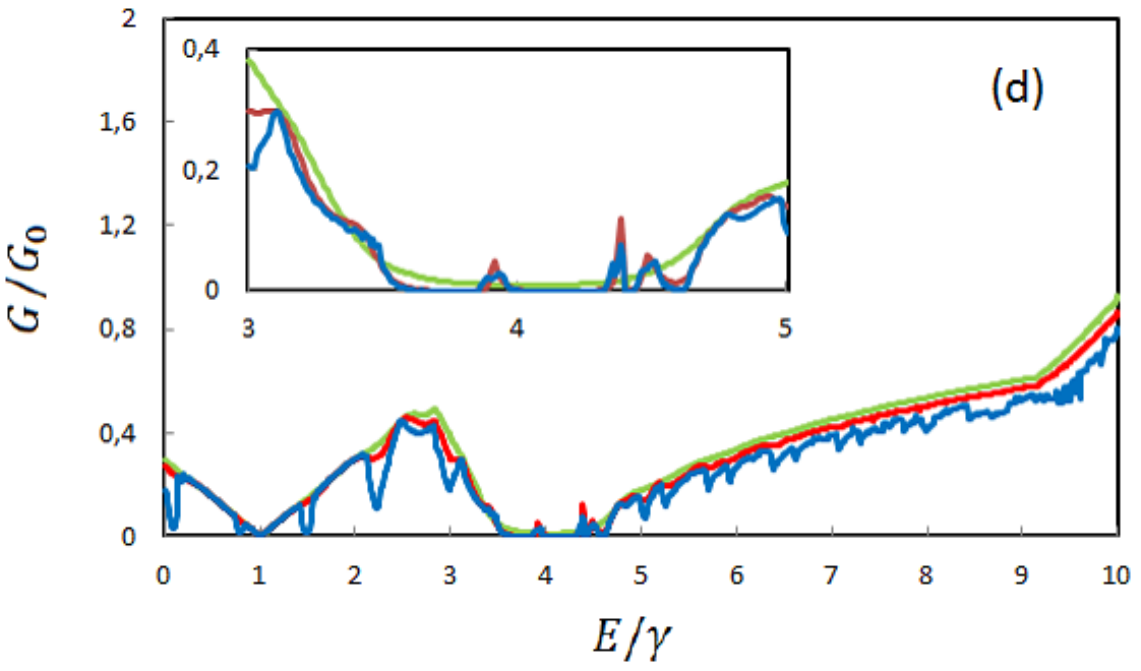}
\caption{\sf Conductance in each individual cone as a function of the
energy for
 $a=2d=30\ nm$ and $V_1=3$, with single (green line) and
 double (red line) barrier structure and also periodic barrier contain $25$ regions (blue line).
 (a)/(c) for $\tau=-1$ and $\Delta_1=0$/{$\Delta_1=0.4\gamma$}. (b)/(d) for $\tau=1$ and
 $\Delta_1=0$/{$\Delta_1=0.4 \gamma$}} \label{fig1.cond}
\end{figure}

{We present in Figure \ref{fig1.cond} the
conductance in each individual cone, given in \eqref{condtau}, for
a single (green line), double barriers (red line) and also for a
finite periodic barrier contain $25$ regions (blue line), as a
function of the energy $E$, for $V_1=3$ and
$a=2d=30\ nm$. 
In Figures \ref{fig1.cond}(a) and \ref{fig1.cond}(b) the
conductances are plotted for $\Delta_{1}=0$ while in Figure
\ref{fig1.cond}(c) and \ref{fig1.cond}(d) for $\Delta_{1}=0.4$.
It is important to note that, in the case of single, double and
finite periodic barriers, we have the same position of the Dirac
point for each individual cone ($E=V_1-1$ in Figure
\ref{fig1.cond}(a) and $E=V_1+1$ in Figure \ref{fig1.cond} (b)).
We conclude that the position of the Dirac point is the same whatever the
number of the barriers. In addition, 
for the case of finite periodic barrier we can have more than one
Dirac point at the same position.
Moreover, when we increase the number of barrier
the conductance decrease and new peaks appear.
To see how the gap will affect the conductance channel in each
cone ($\tau=\pm 1$), we extend }the results presented in Figures
\ref{fig1.cond}(a) and \ref{fig1.cond}(b) to the case
{$\Delta_1=0.4 \gamma$} to get Figures
\ref{fig1.cond}(c) and \ref{fig1.cond}(d). {For the case of single
barrier
 (green line), the
conductance is zero and there are no resonances in the regime of
energy ($V_1+\tau-\Delta_1 < E < V_1+\tau+\Delta_1$). However, for
the double barrier structure (red line) new peaks of resonances
appear in the above mentioned regime of energy.} These peaks can
be attributed to the bound electron states in the well region
between the barriers. The number of these peaks, in the regime of
energy ($V_1+\tau-\Delta_1 < E < V_1+\tau+\Delta_1$), is
the same as that of the finite periodic {barrier.}\\

\begin{figure}[H]
 \centering
  \includegraphics[width=6.5cm, height=4.5cm]{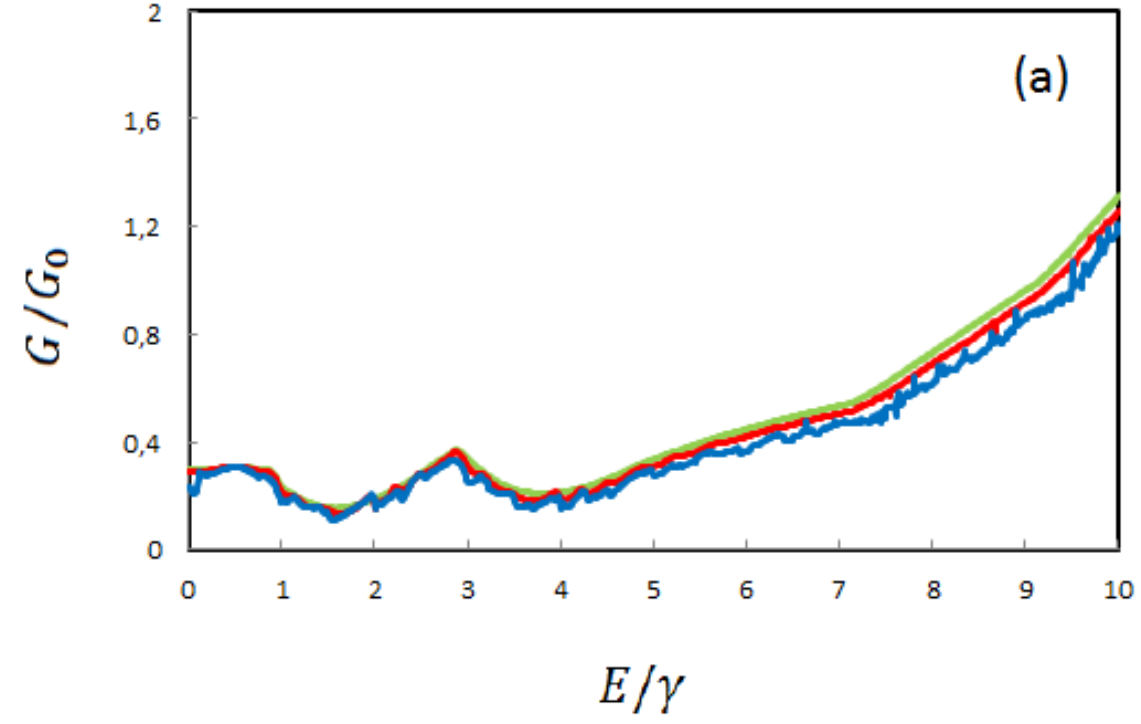}
~~~~~~\includegraphics[width=6.5cm,height=4.5cm]{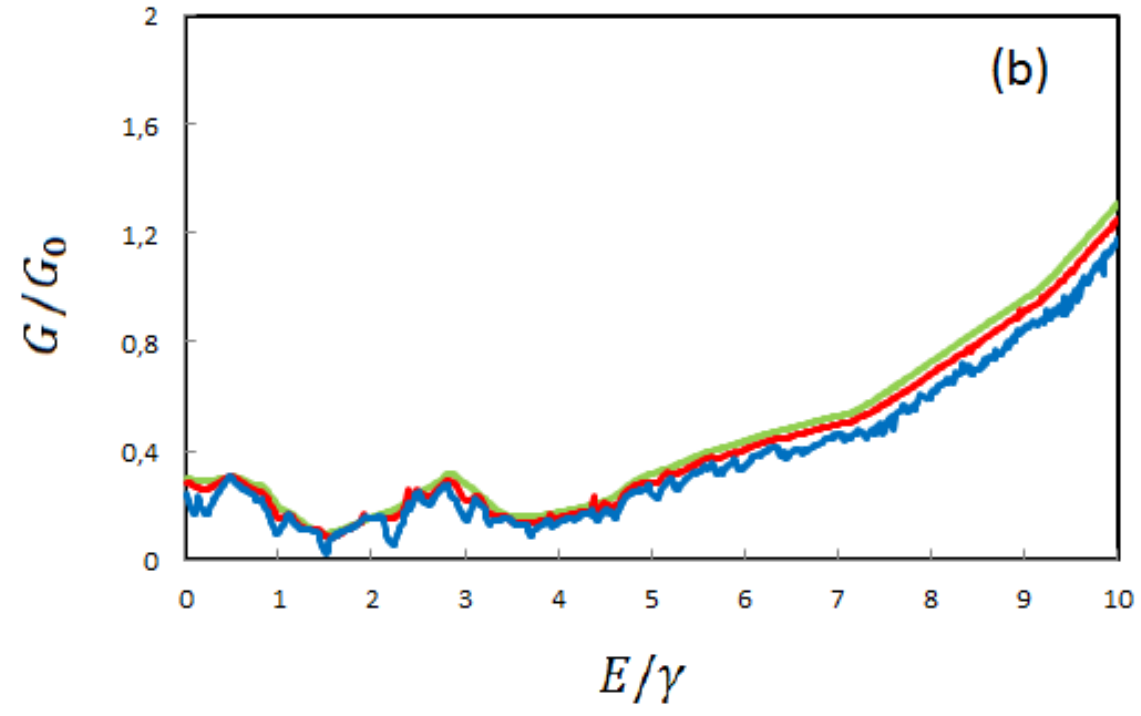}
 \caption{\sf Total conductance as a function of energy, for
$V_1=3$ and $a=2 d=30\ nm$, with single (green line) and
 double (red line) barrier structure and also finite periodic barrier contain $25$ regions (blue line). (a)
 for $\Delta_1=0$ and (b) for
 {$\Delta_1=0.4\gamma$}
 }\label{fig2.condtotal}
\end{figure}

Now, we plot the {total conductance} 
\eqref{condtotal} as a function of energy {for
single (green line), double (red line) barrier structure and for
finite} periodic barrier contain $25$ regions (blue line).
Indeed, the existence of the  intracone
transition allows only two transmission channels in AA-stacked
bilayer graphene \cite{Sanderson} resulting in a total
conductance, while the transmission in the intercone is zero. In
contrast, for AB-stacked bilayer graphene \cite{Duppen} we have
intracone and intercone transitions, which are possible between
all four bands. In addition, the total conductance of AA-stacked bilayer graphene
is different with that of the single layer
graphene \cite{Azarova}.\\

\section{Conclusion}

We have studied the electronic transport of electrons through
single, double and finite periodic barrier on the AA-staked
bilayer graphene. as well as analyzed the transmission probability
and corresponding conductance
we have started by formulating our  Hamiltonian model that describes the
system under consideration and getting  the associated energy bands.
The obtained bands are composed of two Dirac cones shifted up and down by
the interlayer coupling.

Subsequently,  we have calculated the transmission probabilities
and presented numerically the results for each individual cone
($\tau=\pm 1$).
For AA-stacked bilayer graphene the transmission in the intercone
is zero, in contrast, for AB-stacked bilayer graphene where we
have intracone and intercone transitions. By increasing the number
of barriers an occurrence of resonance peaks in the transmission
probabilities. These resonance peaks correspond to the positions
of new cones index like associated with $\tau = \pm 1$ in the
finite periodic barrier structures. However, the location of Dirac
point ($E=V_1+\tau$) does not depend on the number of barriers.
For a gapless graphene we have perfect transmission that is a
manifestation of Klein tunneling. However, the opening gap in the
barrier region leads to suppression of the Klein tunneling effect
for near-normal incidence.

Basing on the obtained results for the transmission probabilities
we have found that there are two conductance channels through different barriers
cases.
In the case of finite periodic barrier we can have more then one
Dirac point at the same position. Also, we have investigated the
effect of the gap on the conductance channel in each cone. It was noticed that in the
well region between the barriers the appearance of  new peaks of resonances, which
can be attributed to the bound electron states. Finally the total
conductance studied as an average of the two conductance channels in
each cone and we have concluded that this later for AA-stacked bilayer graphene
is different with that corresponding to single layer graphene.

\section*{Acknowledgment}

 The generous support provided by the Saudi Center
for Theoretical Physics (SCTP) is highly appreciated by all
authors.

\end{document}